\documentclass[10pt,letterpaper]{article}
\usepackage{opex3}

\begin{document}

\title{Long-distance remote comparison of ultrastable optical frequencies with $ 10^{-15} $ instability in fractions of a second}

\author{A.~Pape$^{1,*}$, O.~Terra$^2$, J.~Friebe$^1$, M.~Riedmann$^1$, T.~W\"ubbena$^1$, E.M.~Rasel$^1$, K.~Predehl $^{2,3}$, T.~Legero$^2$, B.~Lipphardt$^2$, H.~Schnatz$^2$, and G.~Grosche$^2$}

\address{$^1$Institut f\"ur Quantenoptik, Leibniz Universit\"at Hannover, Welfengarten 1, 30167 Hannover, Germany
\\
$^2$Physikalisch-Technische Bundesanstalt, Bundesallee 100, 38116 Braunschweig, Germany
\\
$^3$Max-Planck-Institut f\"ur Quantenoptik, Hans-Kopfermann-Stra{\ss}e 1, 85748 Garching, Germany
\\
}

\email{$^*$pape@iqo.uni-hannover.de}


\begin{abstract} We demonstrate a fully optical, long-distance remote comparison of independent ultrastable optical frequencies reaching a short term stability that is superior to any reported remote comparison of optical frequencies. We use two ultrastable lasers, which are separated by a geographical distance of more than 50 km, and compare them via a 73 km long phase-stabilized fiber in a commercial telecommunication network. The remote characterization spans more than one optical octave and reaches a fractional frequency instability between the independent ultrastable laser systems of $ 3 \times 10^{-15} $ in 0.1~s. The achieved performance at 100~ms represents an improvement by one order of magnitude to any previously reported remote comparison of optical frequencies and enables future remote dissemination of the stability of 100~mHz linewidth lasers within seconds. \end{abstract}





\section{Introduction}

\noindent Optical atomic clocks have surpassed state-of-the art microwave clocks in terms of accuracy and stability and allow targeting fractional inaccuracies of 1 part in $ 10^{18} $ \cite{rosenband2008fra,ludlow2008slc}. A crucial element in achieving this performance are ultrastable lasers which currently limit the short term stability of state-of-the-art optical clocks \cite{ludlow2008slc}. Thus, development of techniques to improve the performance of the ultrastable interrogation lasers is a rapidly evolving field \cite{Webster:PRA2008,Millo:PRA2009}. Already today, ultrastable lasers are demanding and cost-intensive devices, and it is becoming apparent, that future 100~mHz linewidth lasers will even more increase in complexity and costs, making these systems affordable possibly only in a few laboratories. Thus, methods are required, to provide these ultrastable optical frequencies to remote laboratories over large distances with highly demanding short term stability. Direct comparisons of optical clocks or even only state-of-the-art lasers are difficult, since today, the complex setup of an optical clock does not allow for transportation. Since the first demonstrations \cite{Amy-Klein:05}, phase-coherent long-range dissemination and remote characterization of ultrastable optical frequencies have therefore become an important tool in frequency metrology. The most promising method with the highest demonstrated stability directly transmits the ultrastable optical carrier via phase-stabilized fibers \cite{ludlow2008slc,ma1994dso,ye2003dhs,grosche2007toc,foreman2007aip} and was applied to remote optical clock comparisons over distances of 3.5 km \cite{ludlow2008slc,foreman2007cop}. Long-range comparisons recommend usage of commercial telecommunication fiber networks and transfer at 1.5~$ \mathrm{\mu} $m \cite{grosche2007toc,newbury2007,Jiang:08,Grosche2009oft,Kefelian:09,hong2009,Terra:09}. 

\begin{figure}[tb]
\centering\includegraphics[width=8.3cm]{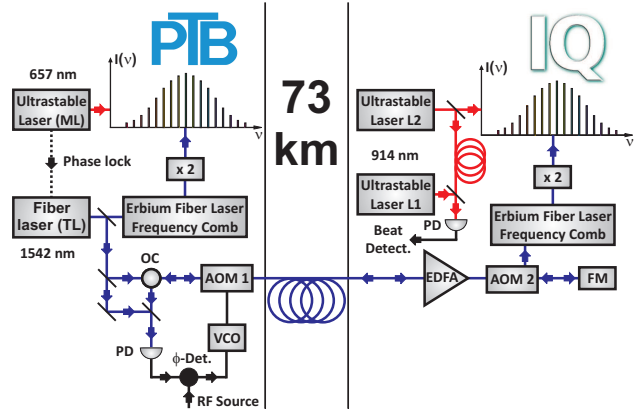}
\caption{Schematic setup. EDFA: bi-directional erbium doped fiber amplifier, AOM: acousto-optic modulator, OC: optical circulator, FM: Faraday mirror, PD: photodiode, \mbox{$ \phi $-Det:} phase detector, VCO: voltage-controlled oscillator.}
\label{Setup}
\end{figure}

In this work we realized a remote comparison of 73~km-distant, independent ultrastable optical frequencies on the $ 10^{-15} $ level in fractions of a second via a long-distance telecommunication fiber network. This represents a remote optical frequency comparison with an unprecedented resolution on short time scales.
The focus of our work was to explore the capability of long-distance fiber links for comparing optical frequencies with high short term stability. Thus, investigating possible limitations on the achievable short term performance of the whole dissemination chain.
The achieved short term stability at 100~ms is a tenfold improvement to previous remote comparisons of optical frequencies. The demonstrated performance allows remote dissemination of the stability of 100~mHz linewidth lasers within seconds. With our setup we are able to reveal the flicker floor of our ultrastable lasers of $ 3 \times 10^{-15} $ in only 0.1~s averaging time. This represents a real-time application of a long-distance fiber link at the performance-level of state-of-the-art optical clocks. In combination with sophisticated fiber stabilization systems \cite{Grosche2009oft}, the results confirm the potential of long-distance fiber links for comparing optical clocks at the level of $ 10^{-17} $ or below in a few minutes.

\section{Setup}

The lasers compared in this work are separated by more than 50~km geographical distance. The individual laser systems are used for a calcium optical frequency standard located at the Physikalisch-Technische Bundesanstalt (PTB) in Braunschweig and for a magnesium optical frequency standard at the Institute of Quantum Optics (IQ) at the University of Hanover. The experimental setup is schematically depicted in Fig. \ref{Setup}. In Hanover, currently two ultrastable laser systems are in operation, L1 and L2. Both are diode lasers systems at 914 nm, which are stabilized to two independent ultrastable optical resonators. L1 uses a horizontal resonator (finesse $ \mathcal{F}=40,000 $) mounted in a multistage spring pendulum configuration for vibration isolation. L2 is stabilized to a horizontal resonator ($ \mathcal{F}=600,000 $) mounted near the symmetry plane for reduced vibrational sensitivity, similar to a design used in \cite{Webster:PRA2007}. These systems are compared to an ultrastable laser located at PTB via a 73~km long telecommunication fiber to IQ. The ultrastable master laser (ML) at PTB is a diode laser system at 657~nm stabilized to a vibrationally insensitive optical resonator reaching a demonstrated linewidth at Hz level \cite{stoehr2006dlh,nazarova2006vir} and a flicker floor of $ \approx 2 \times 10^{-15} $ for $ 0.1-20 $~s \cite{Lipphardt:IEEE2009}. Its stability is transferred to a 1542~nm fiber laser (TL) by means of a femtosecond frequency comb \cite{grosche2008ofs}. We inject about 5~mW of the light of TL into the 73~km long fiber link to IQ, connected via the local computer center in Hanover (RZ-H). At IQ, the light is amplified in a bidirectional erbium doped fiber amplifier (EDFA), frequency shifted with an acousto-optic modulator (AOM 2) and afterwards partially reflected back to PTB, using a Faraday rotator mirror. At PTB, the back reflected light is used for active cancellation of phase noise of the fiber by servo feedback to AOM 1. The servo bandwidth of the noise cancellation loop is limited to $ \approx $~700~Hz due to fiber length \cite{newbury2007}. The single pass loss in the 73~km fiber is $ \approx $~23~dB. Information about the stabilization system and the link between PTB and RZ-H can be found in \cite{Grosche2009oft,Terra:09}. Part of the transmitted light at 1542~nm is coupled out at IQ and used for comparison to L1 and L2 at 914~nm by means of a second frequency comb. Both frequency combs are based on femtosecond erbium fiber lasers which are frequency doubled into the visible. Thus, optical frequencies even far outside from the telecommunication window at 1.5~$ \mathrm{\mu} $m can be remotely compared via the fiber link using the transfer laser at 1542~nm without degrading the stability or accuracy \cite{Grosche2009oft}.

\begin{figure}[tb]
\centering\includegraphics[width=8.3cm, bb = 0 15 301 208]{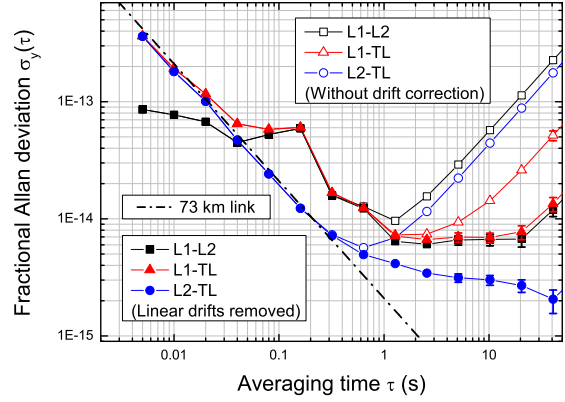}
\caption{Fractional Allan deviation of the beat frequency between ultrastable laser systems in more than 50~km distant laboratories. L1-L2 local measurement, L1-TL and L2-TL via a 73 km stabilized fiber link; TL: transfer laser at 1542~nm. With (filled symbols) and without (open symbols) removal of linear drifts. Also shown, estimated instability of 73 km link from independent round-trip link measurements.}
\label{Stability}
\end{figure}

The remote comparison of the ultrastable optical frequencies is performed in the time and frequency domains. In the time domain, all frequencies are recorded simultaneously by using a multichannel frequency counter with synchronous readout and zero deadtime. This can be operated either with a simple single average per gate time (so called $ \Pi $ estimator) or using overlapping multiple averages within the gate period (overlapping $ \Lambda $ estimator) \cite{dawkins2007IEEE}. The frequency of the transfer beat $ \nu_2 - \frac{m_2}{m_1} \nu_1 $ between the delivered cw light $ \nu_1 $ at 1542 nm and the 914 nm cw light $ \nu_2 $ of the local ultrastable lasers is calculated using the relation
\begin{equation}
\nu_2 - \frac{m_2}{m_1} \nu_1 = \left(2-\frac{m_2}{m_1} \right) \nu_{\mathrm{CEO}} - \frac{m_2}{m_1} \nu_{\mathrm{B1}} + \nu_{\mathrm{B2}},
\label{Transfer}
\end{equation} by measuring the mode numbers $ m_1 $, $ m_2 $ of the frequency comb, the carrier-envelope offset frequency $ \nu_{\mathrm{CEO}} $, and the beat note frequencies $ \nu_{\mathrm{B1}} $, $ \nu_{\mathrm{B2}} $ between the cw frequencies and the corresponding comb modes. $ \nu_1 $ at 1542~nm is thereby measured using the fundamental comb spectrum, $ \nu_1 = m_1\nu_{rep} + \nu_{CEO} + \nu_{B1} $, and $ \nu_2 $ at 914~nm using the frequency doubled comb spectrum, $ \nu_2 = m_2\nu_{rep} + 2\nu_{CEO} + \nu_{B2} $, with $ \nu_{rep} $ the repetition rate of the frequency comb.

\section{Results}

\subsection{Remote 3-laser characterization}

First, we performed a remote comparison between the three independent laser systems. This enabled us to identify the system that provides the best stability for the investigation of the achievable short term performance of the remote comparison.

The results of the remote characterization of the three independent laser systems are depicted in Fig. \ref{Stability}. The Allan deviation (ADEV) shown is obtained by operating the counter with a simple single average per gate time ($ \Pi $ estimator). L2 shows a drift rate of several $ 10^{-15} $/s and L1 of $ 10^{-15} $/s, which we attribute to temperature fluctuations of the ultrastable resonators. The drift rate of ML, and thus TL, was identified independently at PTB using a hydrogen maser as $ 10^{-16}$/s. With linear drift correction, for $\tau \geq 0.05 $~s TL is identified as the most stable laser, and L2 as more stable than L1 (Fig. \ref{Stability}). Also shown in Fig. \ref{Stability} is the link instability of  $ \sigma_y(\tau) = (2.1 \pm 0.5) \times 10^{-15}(\tau/\mathrm{s})^{-1} $, estimated \cite{Williams:08} from independent PTB-IQ-PTB round-trip measurements. According to these estimates, it is reasonable that the observed measurement performance between L2 and TL is limited by the link instability up to 1~s.

\begin{figure}[tb]
\centering\includegraphics[width=8.3cm, bb = 0 15 301 208]{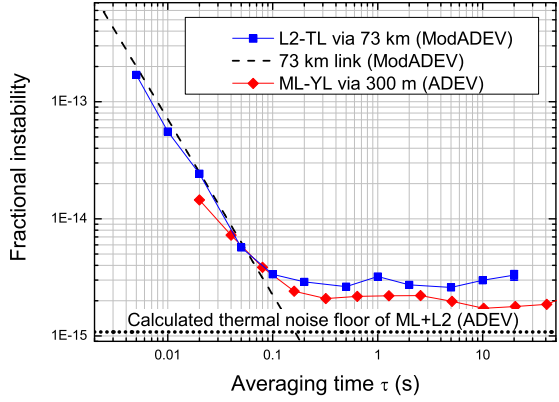}
\caption{High resolution remote optical frequency comparison L2-TL using the modified Allan deviation (ModADEV) and estimated instability of 73 km link from independent round-trip link measurements. Also shown, stability of ML-YL local comparison at PTB via 300~m of stabilized fiber \cite{Lipphardt:IEEE2009}.}
\label{ModAllan}
\end{figure}

\subsection{Rapid high-resolution remote comparison}

To go beyond this limitation and to check for other possible noise sources that are not only introduced by the fiber link itself but also by other components of the whole dissemination chain, as e.g. the two independent frequency combs, we performed phase-averaged data acquisition.

We implemented a data evaluation based on the modified Allan deviation (ModADEV), which is obtained (case (c) in \cite{dawkins2007IEEE}) by operating the counter with overlapping multiple averages per gate time (overlapping $ \Lambda $ estimator, with an internal gate time of 1~ms). While leaving the instability contributions of frequency modulation noise processes unaffected (within 30\% error \cite{dawkins2007IEEE}), the contribution of white phase noise - which is typical for residual link fluctuations - falls off as $ \tau^{-3/2} $. Figure \ref{ModAllan} depicts the ModADEV for the L2-TL comparison. Also shown is the instability of the 73~km link, as estimated from independent round-trip measurements PTB-IQ-PTB. The ModADEV of the link behaves as $ \sigma_y^{\mathrm{mod}}(\tau) = (7.0 \pm 1.8 ) \times 10^{-17} (\tau/\mathrm{s})^{-3/2} $. Thus, at 1~s measurement time, the link instability contribution is reduced by a factor of 30.

The L2-TL trace coincides with the estimated link instability for very short time scales, i.e. no significant additional measurement noise or laser noise could be detected. The performance of the remote laser comparison is already revealed after 100~ms and shows a flicker floor of $ 3 \times 10^{-15} $. The combined thermal noise contributions of the resonators of L2 and ML is calculated \cite{Numata:PRL2009} to be $ 1.1 \times 10^{-15} $. The remote measurement results are close to the performance of state-of-the-art ultrastable lasers, as represented in Fig. \ref{ModAllan} by the previous results of the \textit{local} comparison of ML to the ultrastable interrogation laser YL of the Yb$^+$ experiment at PTB via 300~m of stabilized fiber \cite{Lipphardt:IEEE2009}. The achieved short term stability for the remote comparison is superior to any reported remote comparison of optical frequencies and shows an improvement at 100~ms of one order of magnitude to any published results.

\begin{figure}[tb]
\centering\includegraphics[width=8.3cm, bb = 0 15 301 224]{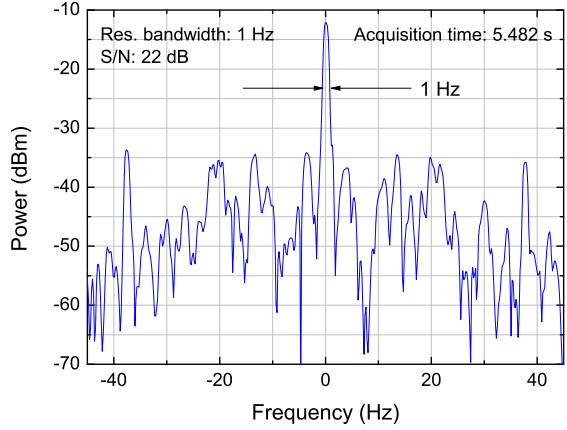}
\caption{Power spectrum of the remote transfer beat note between 73 km distant L2 and TL at 48.5 THz.}
\label{Transferbeat}
\end{figure}

\subsection{Transfer beat between remote laser systems}

For analysis in the frequency domain, we additionally implemented the operation in Eq. (\ref{Transfer}) (divided by 4) by analog signal processing \cite{grosche2008ofs} using a direct digital synthesizer and double-balanced mixers. The spectrum of the remote transfer beat note $ \frac{1}{4}(\nu_2 - \frac{m_2}{m_1}\nu_1) $, at 48.5~THz, between L2 and TL is depicted in Fig. \ref{Transferbeat}. It shows a linewidth of 1~Hz, limited by the spectrum analyzer used. By analyzing the spectrum of the remote beat signal, we optimized the laser system L2 in real-time. The beat signal also enables remote phase locking of a laser to the transferred light at the noise level of the stabilized fiber.

\section{Conclusion}

To conclude, we performed a remote comparison of ultrastable optical frequencies on the $ 10^{-15} $ level on sub-second time scales. ModADEV based data evaluation suppresses residual fiber noise below $ 10^{-16} $ in 1~s and enables rapid resolution of the transferred signal. This displays the enormous potential of dissemination of ultrastable optical frequencies using long-distance telecommunication fiber networks for frequency metrology. Exciting applications are comparisons of optical clocks, remote high-precison spectroscopy or gravitational wave detectors. This work underlines the feasibility of expanding the distances bridged by the optical fiber links. Thus, the vision of a national \cite{Grosche2009oft} and European wide fiber network becomes a realistic scenario in the near future, which will strongly stimulate the field of fundamental physics and precison frequency metrology.

\section*{Acknowledgments}

We gratefully acknowledge long-standing advice and support by W.~Ertmer. We thank T.~Rosenband (NIST, Boulder) for providing us the design for our new ultrastable resonators. We thank U.~Sterr and S.~Kraft for providing the ultrastable calcium clock laser and thank S.~Falke for proof-reading. We acknowledge financial support by the DLR and ESA. O. Terra is supported by a scholarship from Egypt and is a member of the Braunschweig International Graduate School of Metrology, IGSM. This joint work was made possible by the framework of the Sonderforschungsbereich 407 and the Centre for Quantum Engineering and Space Time Research (QUEST), with financial support from the Deutsche Forschungsgemeinschaft.

\end{document}